\documentclass[%
 reprint,
 amsmath,amssymb,
 aps,
 prl,
]{revtex4-1}
\usepackage{graphicx}
\usepackage{dcolumn}
\usepackage{bm}
\usepackage{hyperref}
\usepackage{textgreek}
\usepackage{xcolor}

\begin{document}

\preprint{APS/123-QED}

\title{Temporal cavity solitons and frequency combs via quantum interference}

\author{Gian-Luca \surname{Oppo$^{1}$}}
\email{g.l.oppo@strath.ac.uk}
\author{David \surname{Grant$^{1}$}}
\author{Mansour \surname{Eslami$^{2}$}}
\affiliation{$^1$SUPA and Department of Physics, University of Strathclyde, 
Glasgow, G4 0NG, Scotland, UK\\ 
$^2$Department of Physics, University of Guilan, P.O. Box 41335-1914 Rasht, Iran}

\begin{abstract}
Temporal cavity solitons in ring microresonators provide broad and controllable generation of frequency combs with applications in frequency standards and precise atomic clocks. Three level media in the $\Lambda$ configuration inside microresonators displaying electromagnetically induced transparency can be used for the generation of temporal cavity solitons and frequency combs in the presence of anomalous dispersion and two external driving fields close to resonance. Here, domain walls separating regions of two dark states due to quantum interference correspond to realizations of stimulated Raman adiabatic passage without input pulses. With no need of modulational instabilities, bright temporal cavity solitons and frequency combs are formed when these domain walls lock with each other. Wide stability ranges, close to resonance operation and optimal shape of the cavity solitons due to three-level quantum interference can make them preferable to those in two-level media.
\end{abstract}

\maketitle


Frequency combs for precision optical synthesizing have come a long way in the past two decades from mode--locked lasers with cavity lengths between 30 cm to 3 m to chip-scale sources based on microresonators \cite{Pasquazi18,Fortier19}. When operating at low-loss, microresonators provide long interaction times and can excite large nonlinearities. A considerable attention was paid to the theoretical and experimental development of frequency combs in high-Q optical resonators operating away from material resonances via Kerr nonlinearities. Temporal cavity solitons (TCS) \cite{Leo10}, the analogue of transverse (diffraction) cavity solitons \cite{Scroggie94,Ackemann09} in the longitudinal (dispersion) domain, are key nonlinear solutions for the generation of broad frequency combs in passive \cite{Herr14,Kippenberg18} and active \cite{Bao19} media. These devices benefit from a variety of technological merits including low-energy requirements and robust structures that can be integrated on a chip \cite{Chembo16,Pasquazi18}.

At the same time, substantial progress has been made in the realm of quantum interference phenomena for the manipulation of the optical response of a medium close to material resonances. Electromagnetically induced transparency (EIT) is one such phenomenon where the optical response of the material can be controlled by external electromagnetic fields to forbid, for example, absorption or to enhance the refractive index \cite{Fleischhauer05}. In this transparency window, dispersion properties are also strongly modified motivating many applications such as slow light and stored light \cite{Novikova12}. The underlying mechanism is the quantum interference effect that destructively couples the transition amplitudes of different excitation pathways \cite{Fleischhauer05}. Since then, mircoresonators have become an interesting platform for the realization of EIT leading to solid state based technologies and new applications in sensing and field enhancement \cite{Xiao09,Liu17}. Recently, EIT resonance line shapes have been found in a variety of optical cavities \cite{Qin20} including micro-ring resonators in silicon-on-insulator chips with air holes \cite{Gu18}. On the theoretical side, optical cavities with three-level media displaying EIT have also been studied for the onset and stability of various transverse structures, from patterns to rogue waves to diffractive cavity solitons \cite{Oppo10}.

Yet to be studied are the fast time EIT features for microresonators in the presence of longitudinal dispersion. It is the aim of this communication to show that quantum interference coupled with anomalous dispersion in microresonators displaying EIT can lead to bistability of dark states, stable domain walls (DW) between these states, trapped stimulated Raman adiabatic passage (STIRAP) \cite{Vitanov17} schemes with no input pulses, novel TCS and frequency combs induced by quantum interference and EIT.

\begin{figure}[tbp]
	\begin{center}
		\includegraphics[scale=1.3]{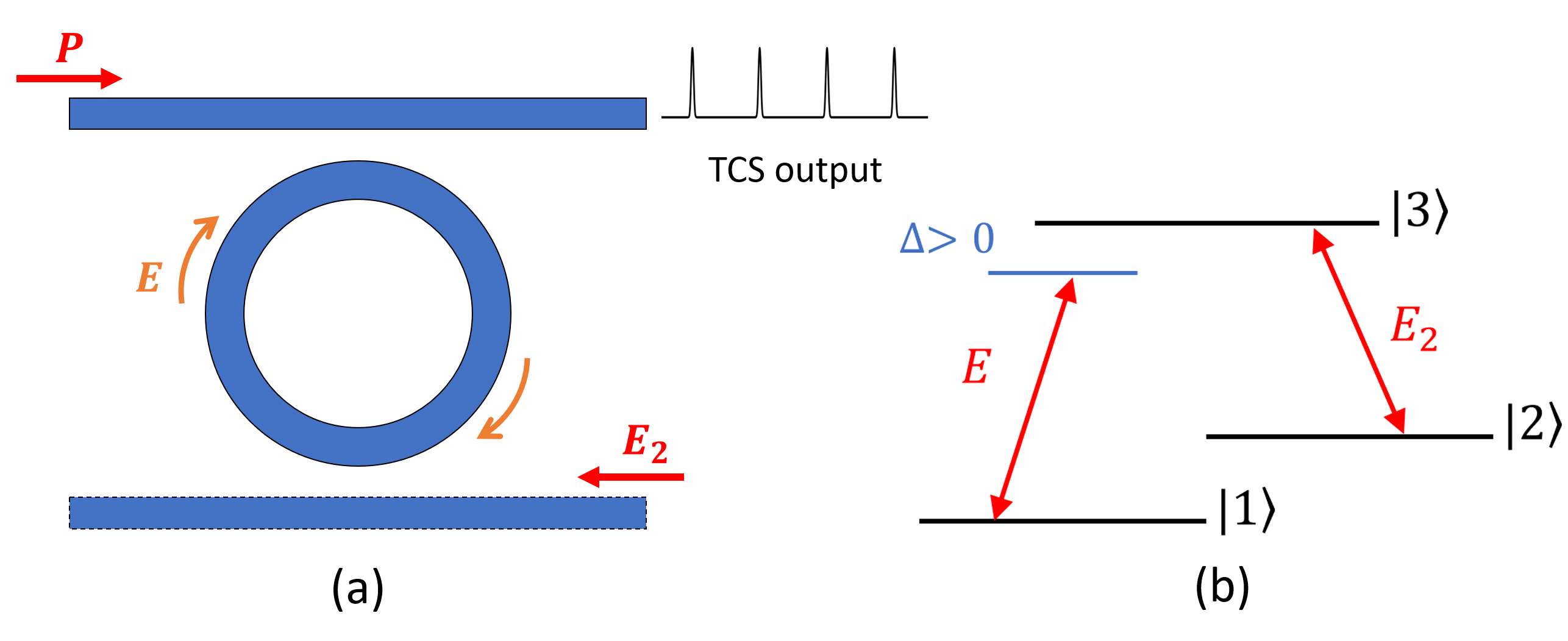}
	\end{center}
	\caption{a) Schematic representation of the dispersive ring-resonator device described by Eq.~\ref{model}.
	b) $\Lambda$ type three energy level configuration with self-focusing nonlinearity.}
	\label{config}
\end{figure}

We consider a ring microresonator as displayed in Fig. \ref{config}a filled with a three level medium in the $\Lambda$ configuration (Fig. \ref{config}b) with the transition between the two lower levels $|1\rangle$ and $|2\rangle$ not dipole allowed. There are two optical beams interacting with the medium, the resonated field $E$ circulating in the ring cavity under the external cw driving $P$ and the non-resonated field of complex amplitude $E_2$ at resonance with the transition between levels $|2\rangle$ and $|3\rangle$. The field $E$ is considered here to be red detuned by $\Delta$ from the resonance of levels $|1\rangle$ and $|3\rangle$. Note that the results presented below extend to wide ranges of $\Delta$, to fields $E_2$ away from resonance and slow relaxations of the level $|2\rangle$ to level $|1\rangle$.
In the case of a fast response medium, the dynamics of the field $E$ in the ring resonator is described by \cite{Haelterman92}:
\begin{equation}
\partial_t E = P-(1+i\theta) E - i (2C) R_{13} + i \partial_{\tau^2} E
\label{model}
\end{equation}
where $t$ is the slow time over several round-trips in the cavity, $P$ is the amplitude of the input pump, $\theta$ is the cavity detuning from the input frequency and $2C$ is the cooperativity parameter of the light-matter coupling that is proportional to the dipole moment of the transition between levels $|1\rangle$ and $|3\rangle$. In \cite{Oppo10} we focused on the transverse diffractive case while here we investigate the anomalous group-dispersion case with a longitudinal variable $\tau$, the fast time in the cavity. $R_{13}$ is the density matrix element in the Lindblad master equation given by \cite{Manka91,Oppo10}: 
\begin{eqnarray}
\label{complete}
 R_{13} &=& \chi(|E|^2) E = \frac{\Delta |E_2|^2 (|E_2|^2+|E|^2-\Delta^2-i\Delta)}{D} \, E \nonumber \\
 D &=& (|E_2|^2+|E|^2 )^3 \\
 &+& \Delta^2 (|E_2|^2+|E|^2+4|E_2|^2|E|^2+\Delta^2 |E_2|^2-2|E_2|^4) \; .\nonumber
\end{eqnarray} 
The detuning $\Delta$ can be scanned by changing the frequency of the input laser. This results in changes of the cavity detuning $\theta$ as well. It is possible, however, to scan the cavity detuning $\theta$ without affecting $\Delta$ by operating on cavity features such as its length. This is different from the two-level medium case where the medium response is basically unaffected by the changes in the frequency of the input beam because one is operating far from the medium resonance. In the following we discuss cavity frequency scans where the input frequency and consequently $\Delta$ are kept fixed.

\begin{figure}[tbp]
	\begin{center}
		\includegraphics[scale=.15]{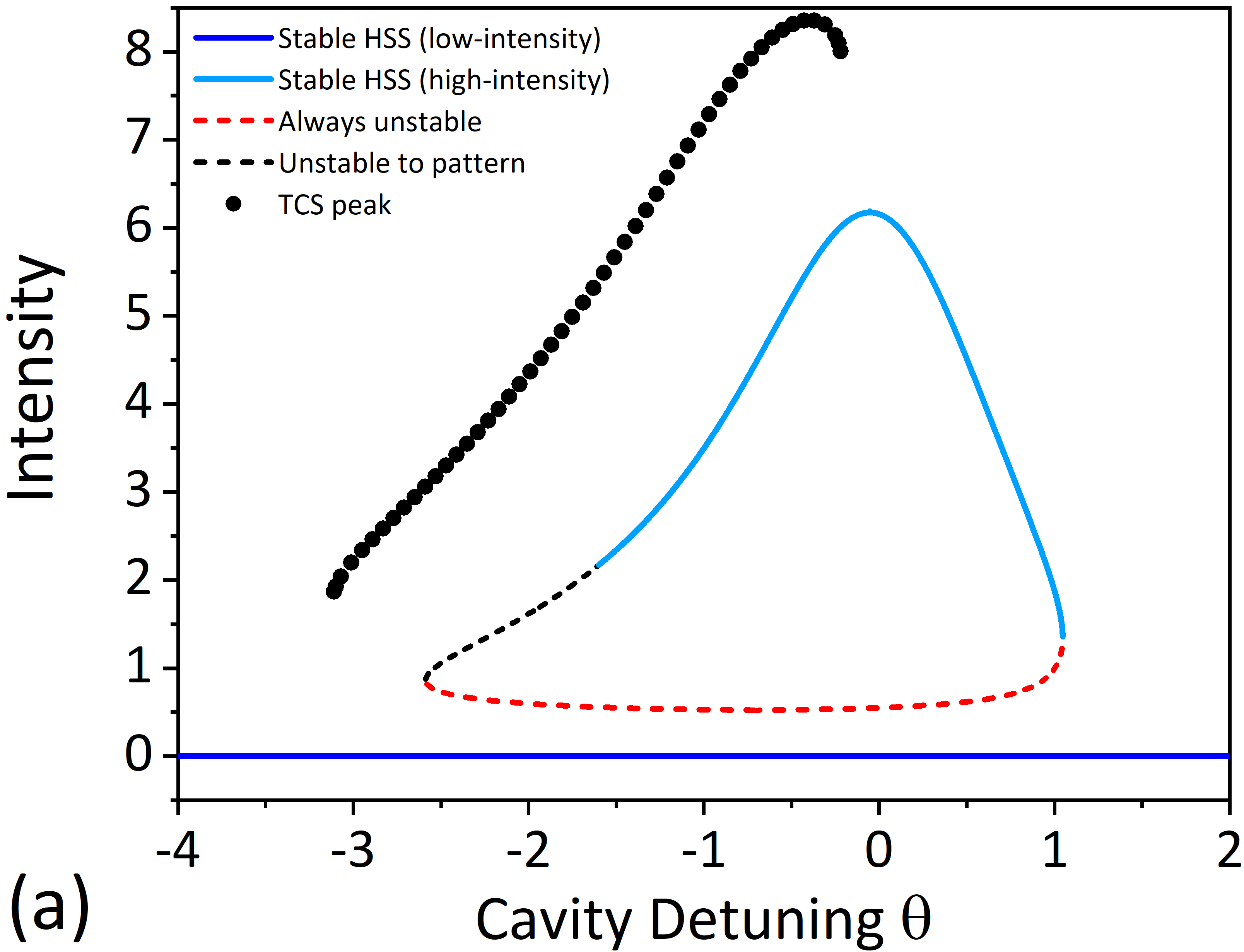}\quad
		\includegraphics[scale=.15]{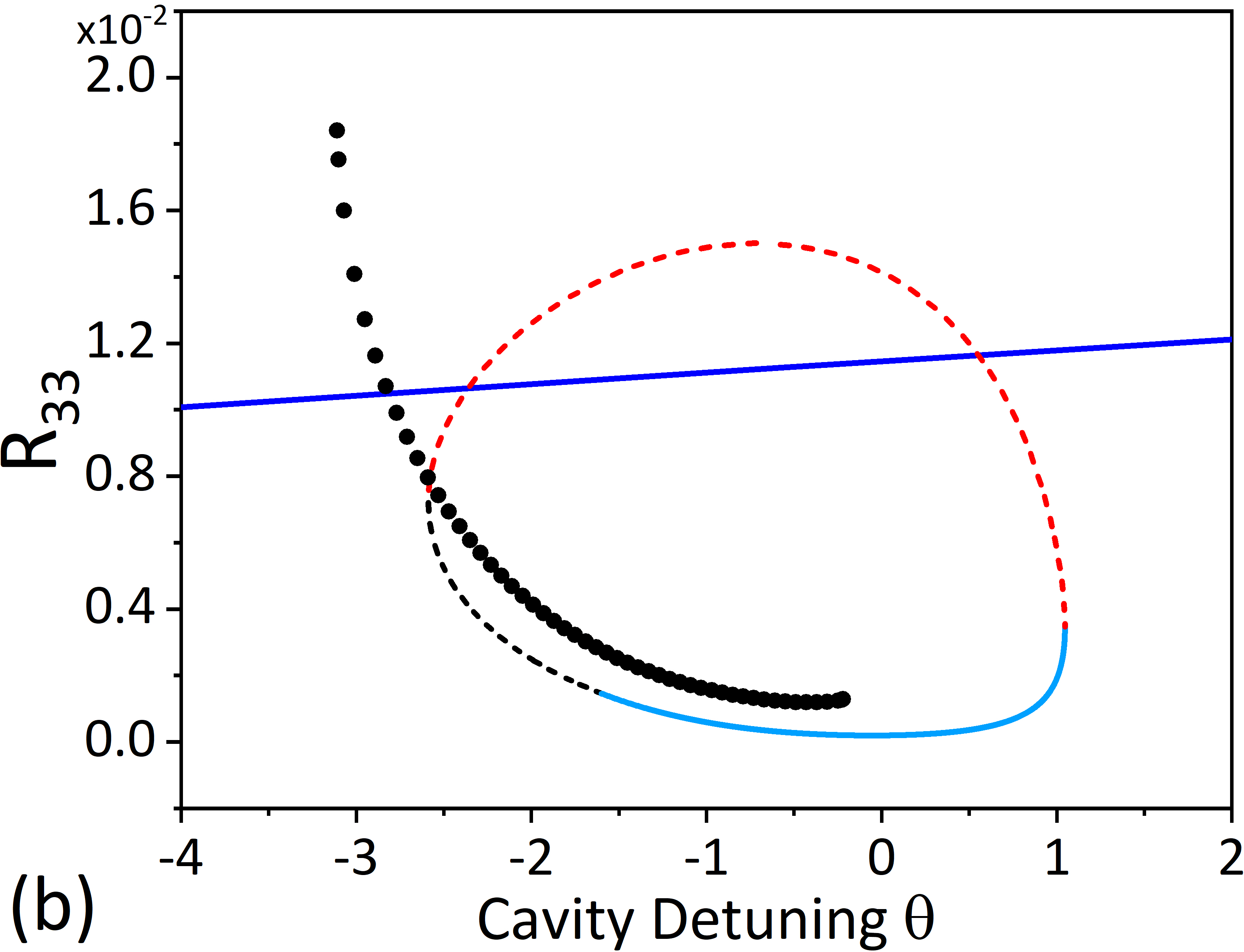}\vspace{.2cm}
    	\includegraphics[scale=.15]{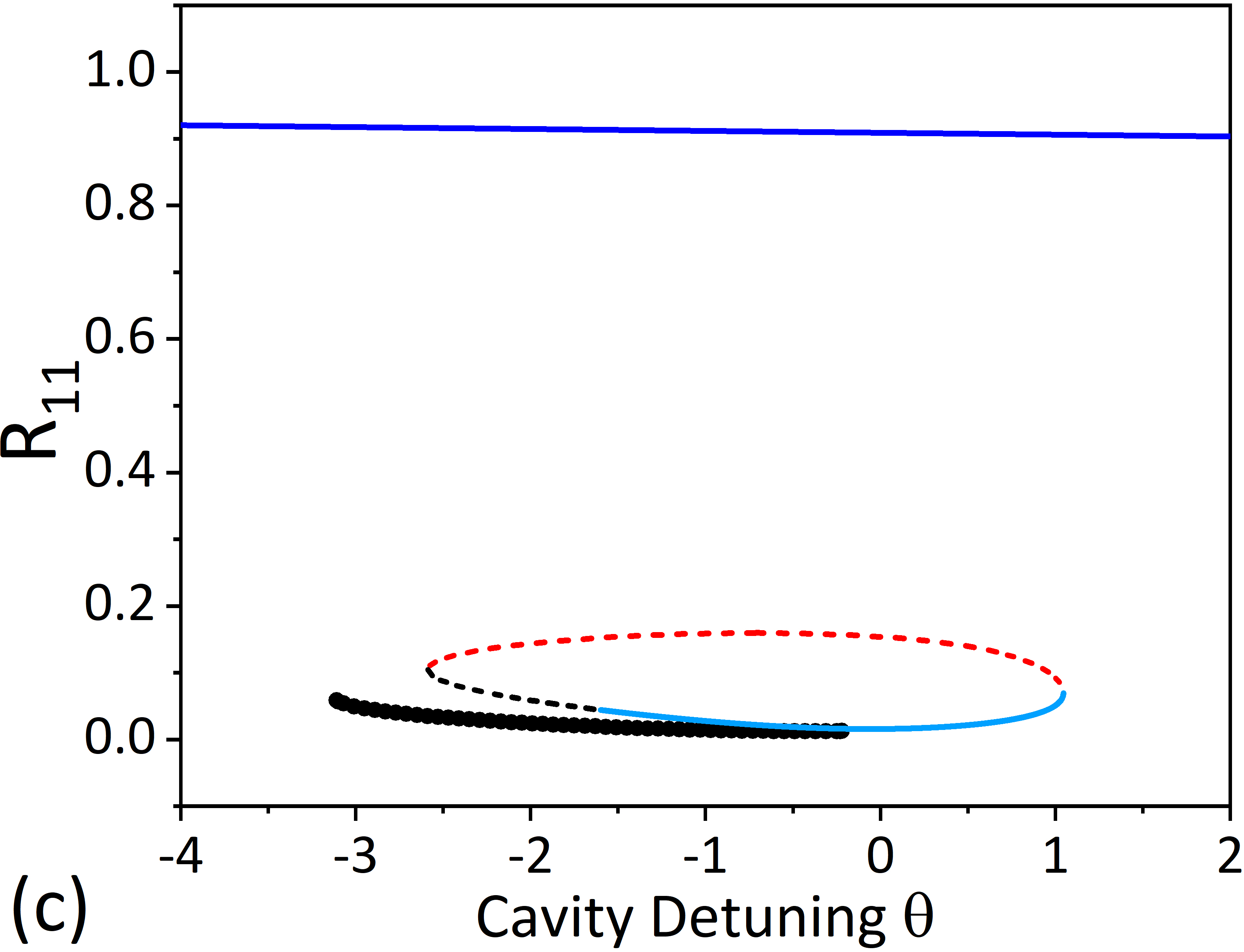}\quad
		\includegraphics[scale=.15]{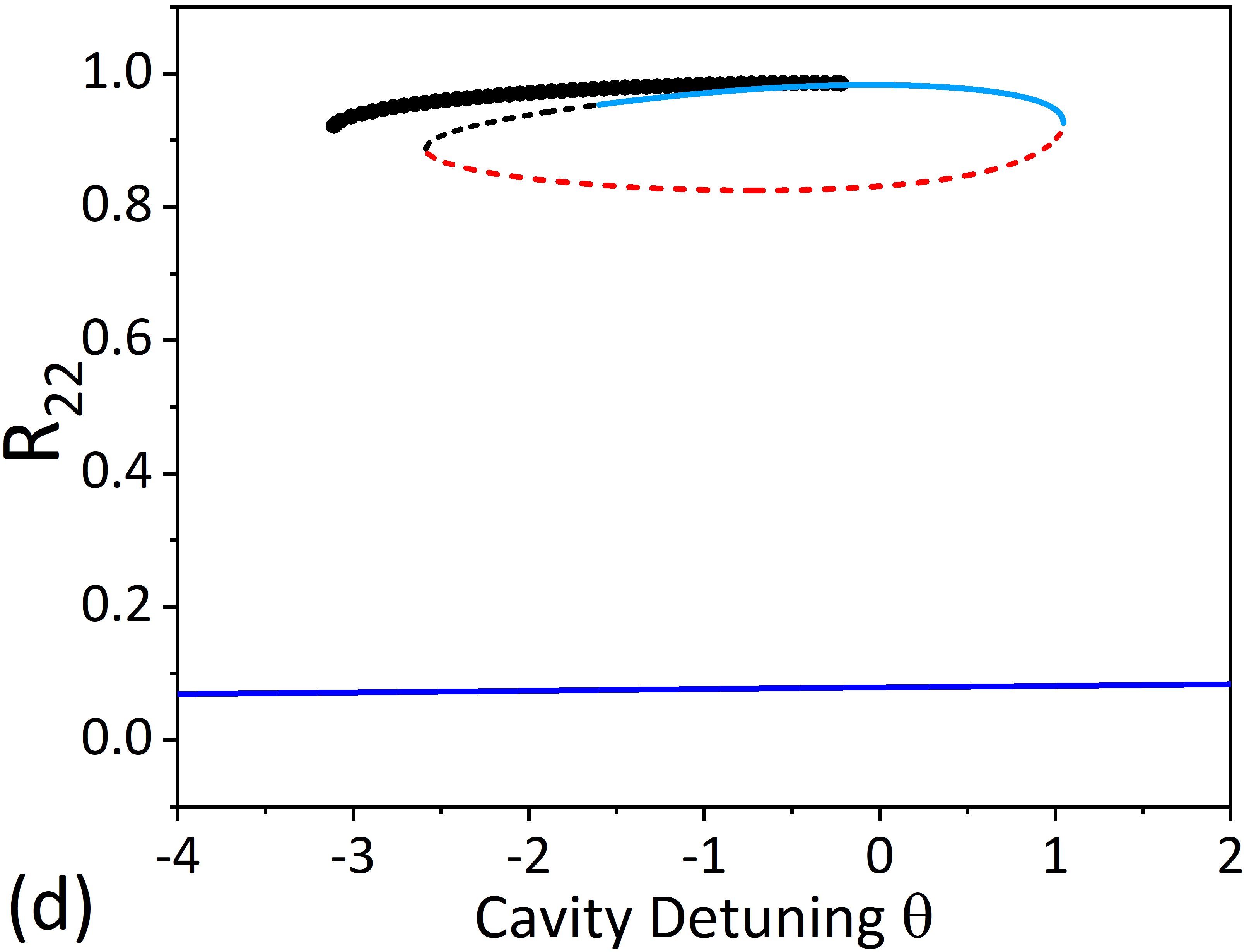}
	\end{center}
	\caption{Intensity (a), $R_{33}$ (b), $R_{11}$ (c) and $R_{22}$ (d) for the HSS of Eq. (\ref{model}) for $|E_2|^2=0.1$, $P=2.5$, $2C=35$, $\Delta=0.66$ when changing the cavity detuning $\theta$. Light and dark blue solid lines are for stable HSS, red and black dashed lines are for unstable HSS, black dots are for TCS peaks.}
	\label{fig2}
\end{figure}

The intensities $|E_s|^2$ of the homogeneous stationary solutions (HSS) are obtained from the equation:
\begin{equation}
\label{HSS}
|P|^2 = \left[ \left( 1- 2C\,{\rm Im}(\chi)\right)^2+\left(\theta+2C\,{\rm Re}(\chi)\right)^2 \right] \, |E_s|^2
\end{equation}
where ${\rm Re}(\chi)$ and ${\rm Im}(\chi)$ are the real and imaginary part of the complex susceptibility of Eq. (\ref{complete}). For small values of the intensity of the second field $E_2$, one observes a typical bistability close to cavity resonance which is, however, strongly enhanced by the quantum interference and splits into a closed bubble and a low-intensity branch as shown in Fig. \ref{fig2}a. 
It is possible to obtain the stability of the HSS through standard linearization techniques. In Fig. \ref{fig2}a for $|E_2|^2=0.1$, $P=2.5$, $2C=35$ and $\Delta=0.66$ we see that the low-intensity HSS branch is always stable, the lower part of the bubble is always unstable to homogeneous perturbations (red dashed curve) while the high-intensity HSS branch is stable for $\theta>-1.6$ and unstable to modulated perturbations (patterns) for $-2.6<\theta<-1.6$. The HSS expressions of the probability of occupancy of each of the three levels are given by:
\begin{eqnarray}
\label{RII}
R_{11} &=& 1-R_{22}-R_{33} \nonumber \\
R_{22} &=& \frac{|E_s|^2 \left[ (|E_s|^2+|E_2|^2)^2+\Delta^2(1+|E_2|^2) \right]}{D} \\
R_{33} &=& \frac{2\Delta^2 |E_s|^2 |E_2|^2}{D} \nonumber
\end{eqnarray}
where $D$ is the denominator appearing in Eq. (\ref{complete}) with the HSS intensities replacing $|E|^2$, and are plotted in Fig. \ref{fig2}b-d. The first key observation is that for generic medium and cavity configurations close to medium resonance, coherent population trapping (CPT) and extremely low values of the probability of occupancy of the excited state $|3\rangle$ are observed (see values of $R_{33}$ below $2\%$ in Fig. \ref{fig2}b). From Fig. \ref{fig2}c-d, we see that the stable low (high) intensity HSS corresponds to maintaining more than $90\%$ of the population in level $|1\rangle$ (level $|2\rangle$). This is a remarkable resonant coupling between the medium properties and the cavity confinement induced by the quantum interference in the three level medium leading to the almost exclusive population of dark states ${\cal D}=\cos(\beta) |1\rangle - \sin(\beta) |2\rangle$ where $\beta$ is the mixing angle with values close to either zero or $\pi/2$. 

\begin{figure}[tbp]
	\begin{center}
		\includegraphics[scale=.15]{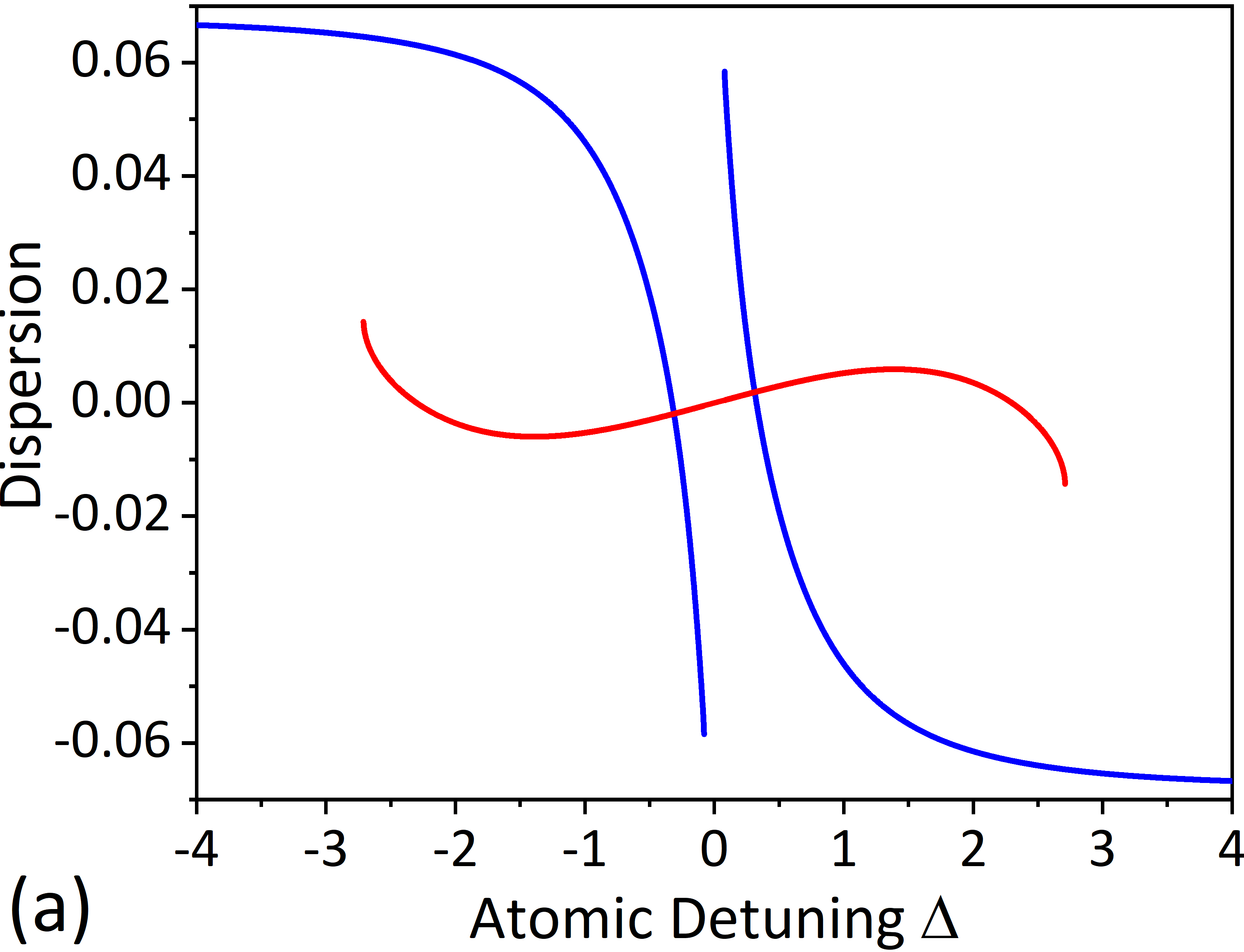}\quad
		\includegraphics[scale=.15]{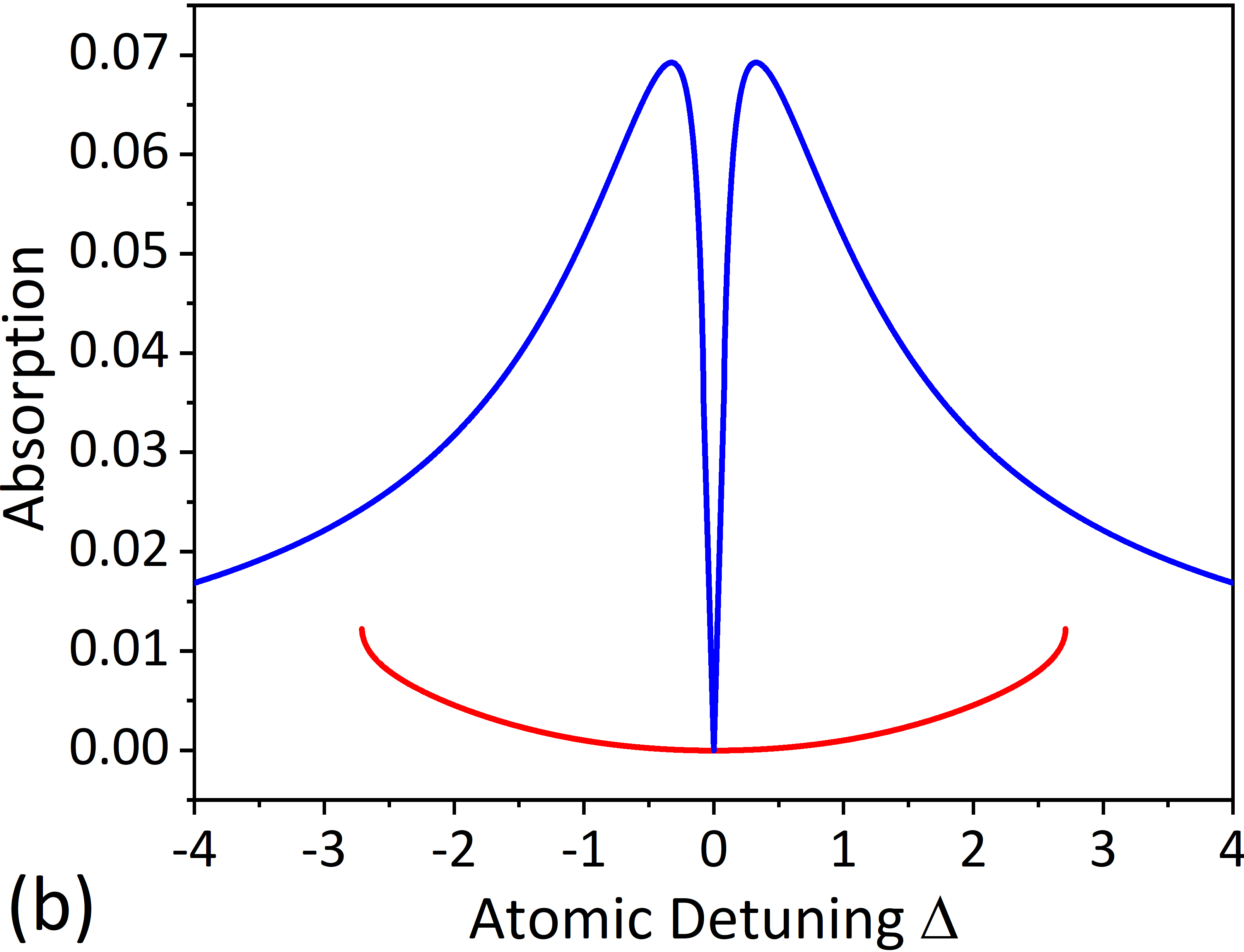}
	\end{center}
	\caption{Chromatic dispersion ${\rm Re}(R_{13})$ (a) and absorption ${\rm Im}(R_{13})$ (b) versus the material detuning $\Delta$ for the $|1\rangle$ -- $|3\rangle$ transition for $\theta=0$, $|E_2|^2=0.1$, and $P=2.5$. Solid blue (red) corresponds to low (high) intensity branch of the bistable HSS.}
	\label{fig3}
\end{figure}

To understand the mechanisms behind these effects, we have plotted in Fig. \ref{fig3} the chromatic dispersion and absorption properties of the transition $|1\rangle$--$|3\rangle$ of the $\Lambda$ system in the two bistable HSS at cavity resonance when changing the detuning $\Delta$.
In the low-intensity branch, chromatic dispersion shifts the input frequency and removes the effect of the cavity resonance. The absorption has a typical EIT shape and is ineffective over a broad band of detunings $\Delta$ thus explaining why $90\%$ of the population remains confined to level $|1\rangle$. The high-intensity branch has almost zero dispersion, is strongly affected by the cavity resonance (note the peak in Fig. \ref{fig2}a close to $\theta=0$) and displays a broad band EIT due to quantum interference. 

\begin{figure}[htbp]
\centering
\includegraphics[scale=.15]{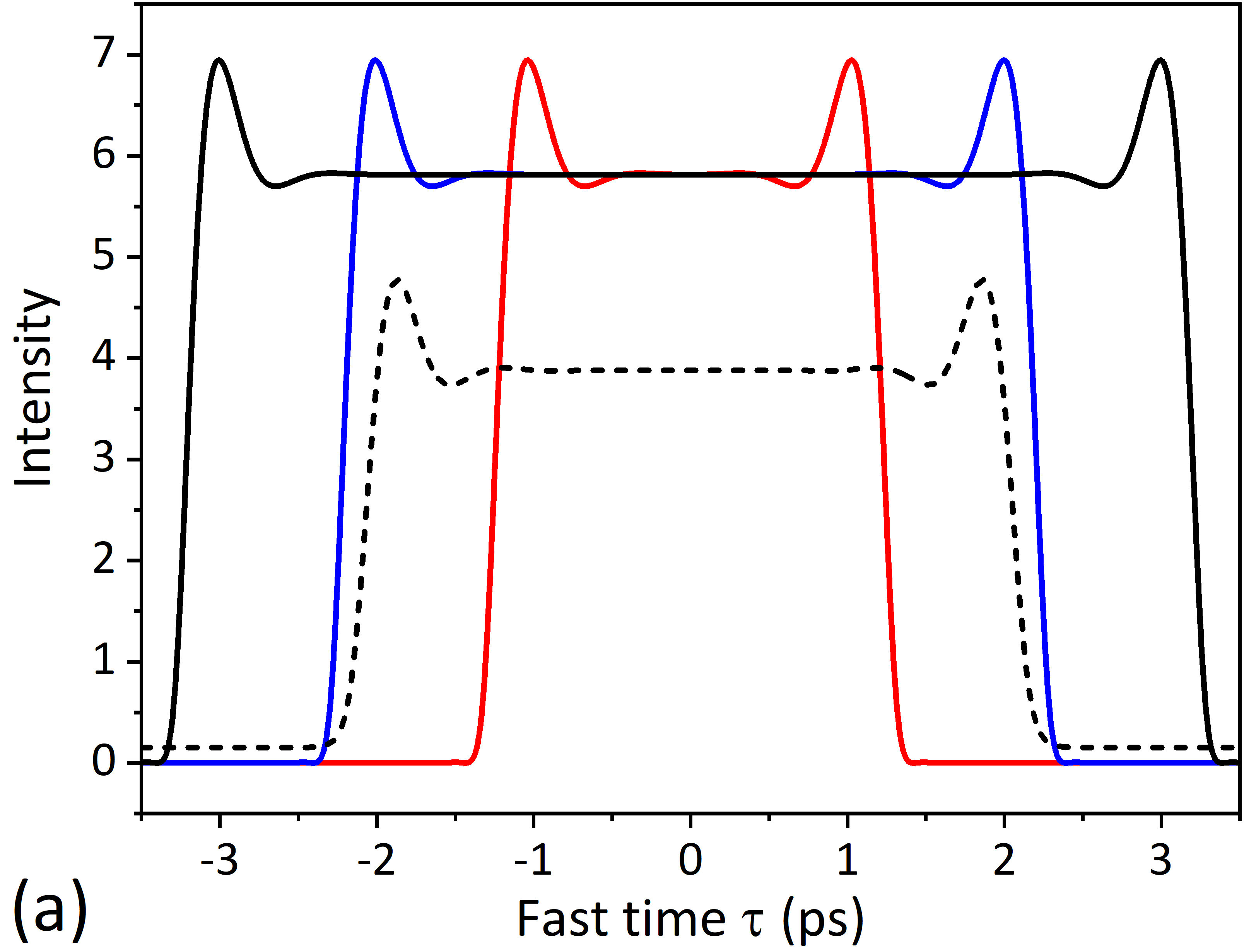}\quad
\includegraphics[scale=.15]{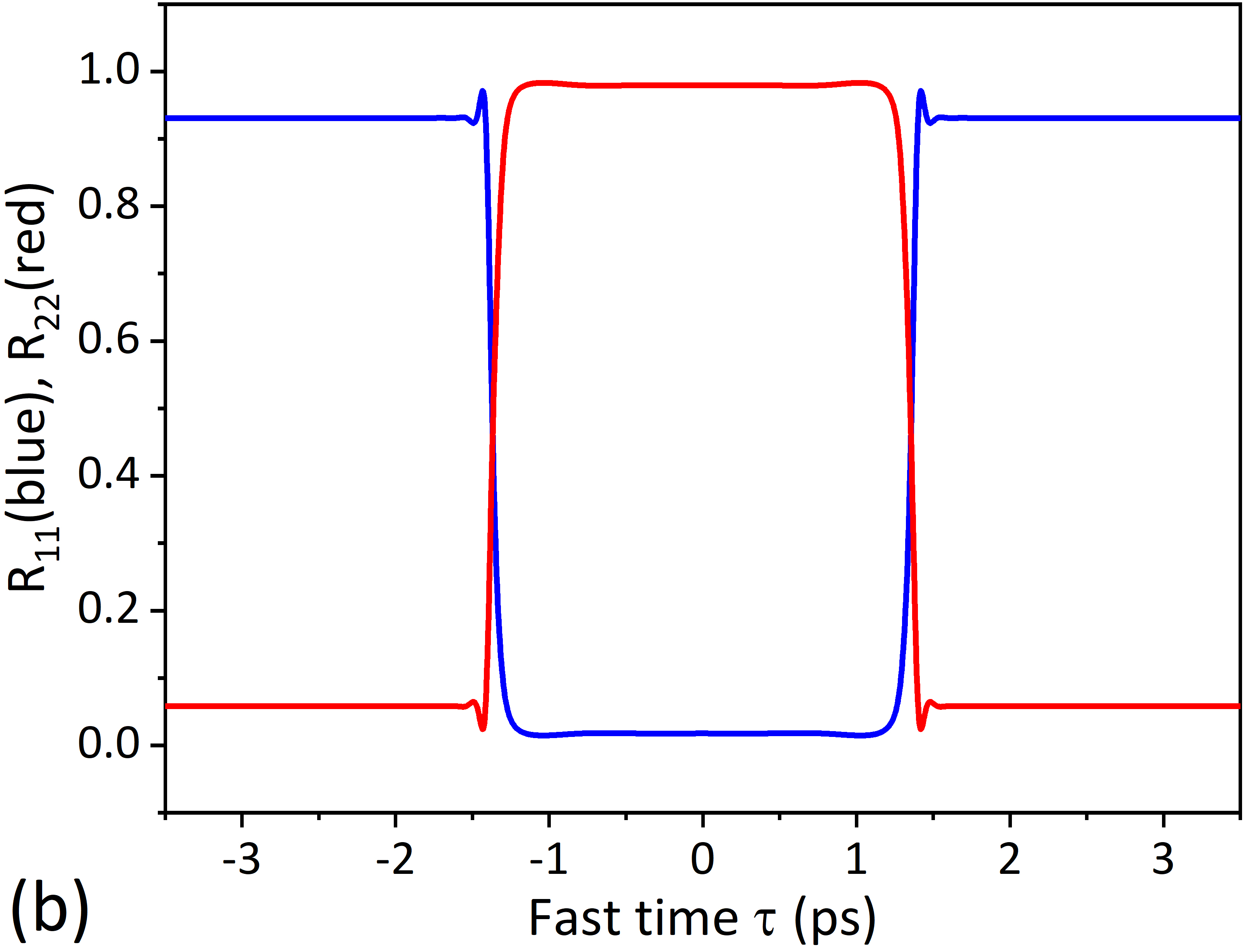}
\caption{a) Intensity profiles of two DWs at stable distances for $\theta_M=-0.305$, $|E_2|^2=0.1$, $P=2.5$, $2C=35$, $\Delta=0.66$ (solid lines) and for $\theta_M=-0.37$, $|E_2|^2=1.0$, $P=2.0$, $2C=35$ and $\Delta=0.2$ (dashed line). b) Fast time distributions of occupancy probabilities of level $|1\rangle$, $R_{11}$ (blue line), and level $|2\rangle$, $R_{22}$ (red line) for $\theta_M=-0.305$, $|E_2|^2=0.1$, $P=2.5$, $2C=35$, $\Delta=0.66$ for the red line profile of a).}
\label{fig4}
\end{figure}

On both sides of cavity resonance, we observe bistability of the low and high intensity branches of the HSS. We have then investigated existence and stability of solutions that move from the low-intensity branch to the high-intensity branch, and vice-versa, during a round trip of the cavity. Steep steps from one branch to the other are known as domain walls (DW) (or switching waves when in motion \cite{Rosanov02}) in analogy with magnetic systems. In nonlinear optics, DWs have been described in optical parametric oscillators \cite{Trillo97,Oppo99} and, more recently, in Kerr resonators in normal dispersion \cite{Xue15,Parra16,Garbin17} and with two orthogonal polarizations \cite{Gilles17,Garbin21}. By tuning $\theta$ it is possible to find Maxwell points, labeled $\theta_M$, where upward and downward DWs exist and are stationary at many separation distances as shown for example in Fig. \ref{fig4} where we have considered microresonators with a free spectral range of around $140$ GHz in agreement with realizations of silicon-on-insulators devices \cite{Yu16}.\\

When moving across a DW inside the resonator, the probability of occupancy goes from being in the state $|1\rangle$ (low output intensity) to state $|2\rangle$ (high output intensity) as displayed in Fig. \ref{fig4}b, while the population of level $|3\rangle$ remains well below $1\%$ due to CPT and EIT. More than $90\%$ of the population of level $|1\rangle$ can be transferred to level $|2\rangle$ when moving from one side of the DW to the other. This is nothing else than a cavity enhanced STIRAP process, the powerful method for efficient and selective transfer of population between two quantum states \cite{Vitanov17}. STIRAP has an enormous number of applications \cite{Bergmann19} and is generally based on the sequence of two gated pulses of laser light with appropriate Rabi frequency (i.e. intensity) profiles. The DW based STIRAP presented here requires no pulsed light at the input. Once the DWs are formed by an initial perturbation, they keep circulating in the resonator with one part of the resonator residing in one of the two (dark) quantum states leaving the other part in the second of the two (dark) quantum states. It is interesting to note that stable DWs correspond to a continuous wave realization of STIRAP since the input pumps $P$ and $E_2$ are not pulsed as it is standard in STIRAP configurations \cite{Vitanov17,Bergmann19}. 

\begin{figure}[htbp]
\centering
\includegraphics[scale=.15]{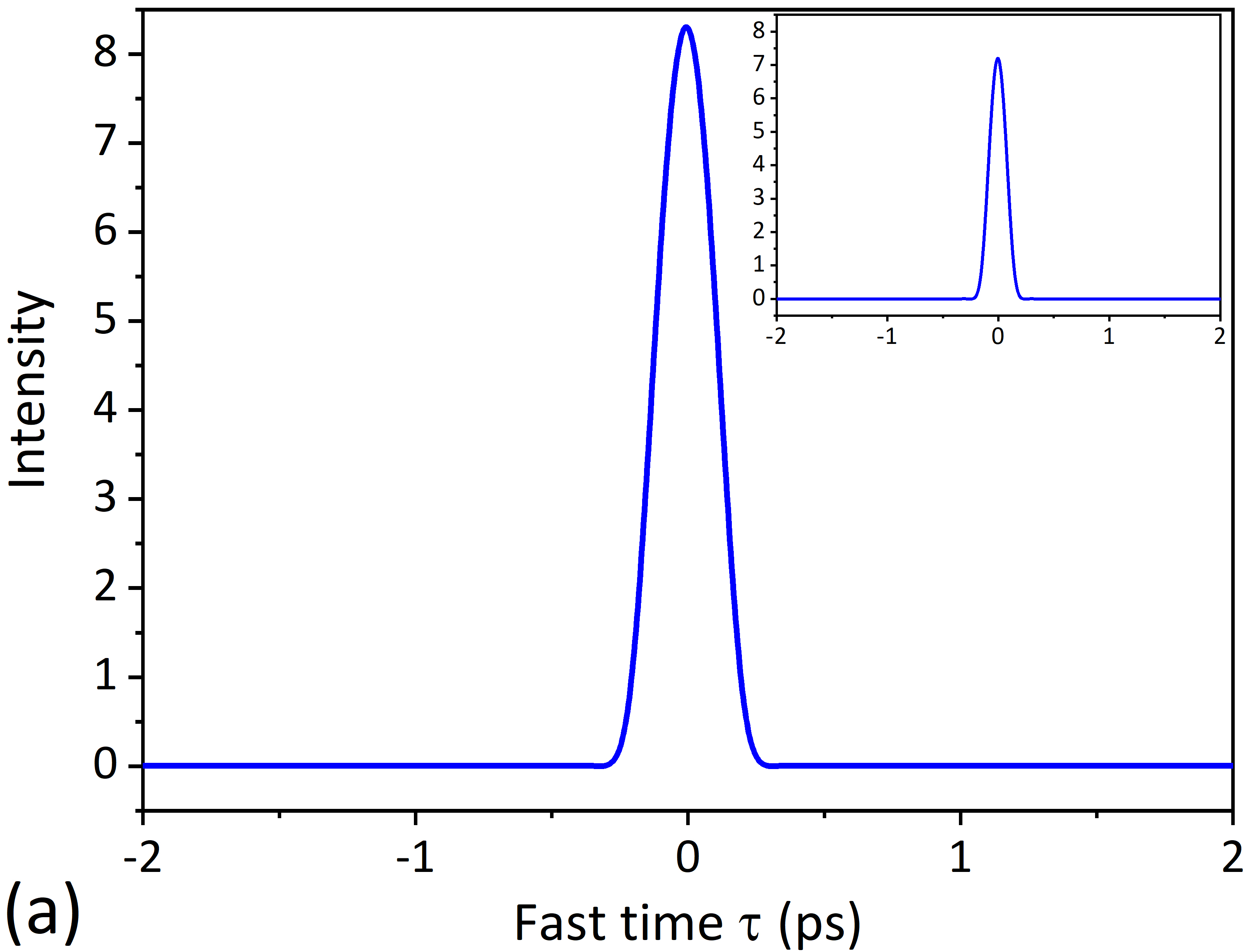}\quad
\includegraphics[scale=.15]{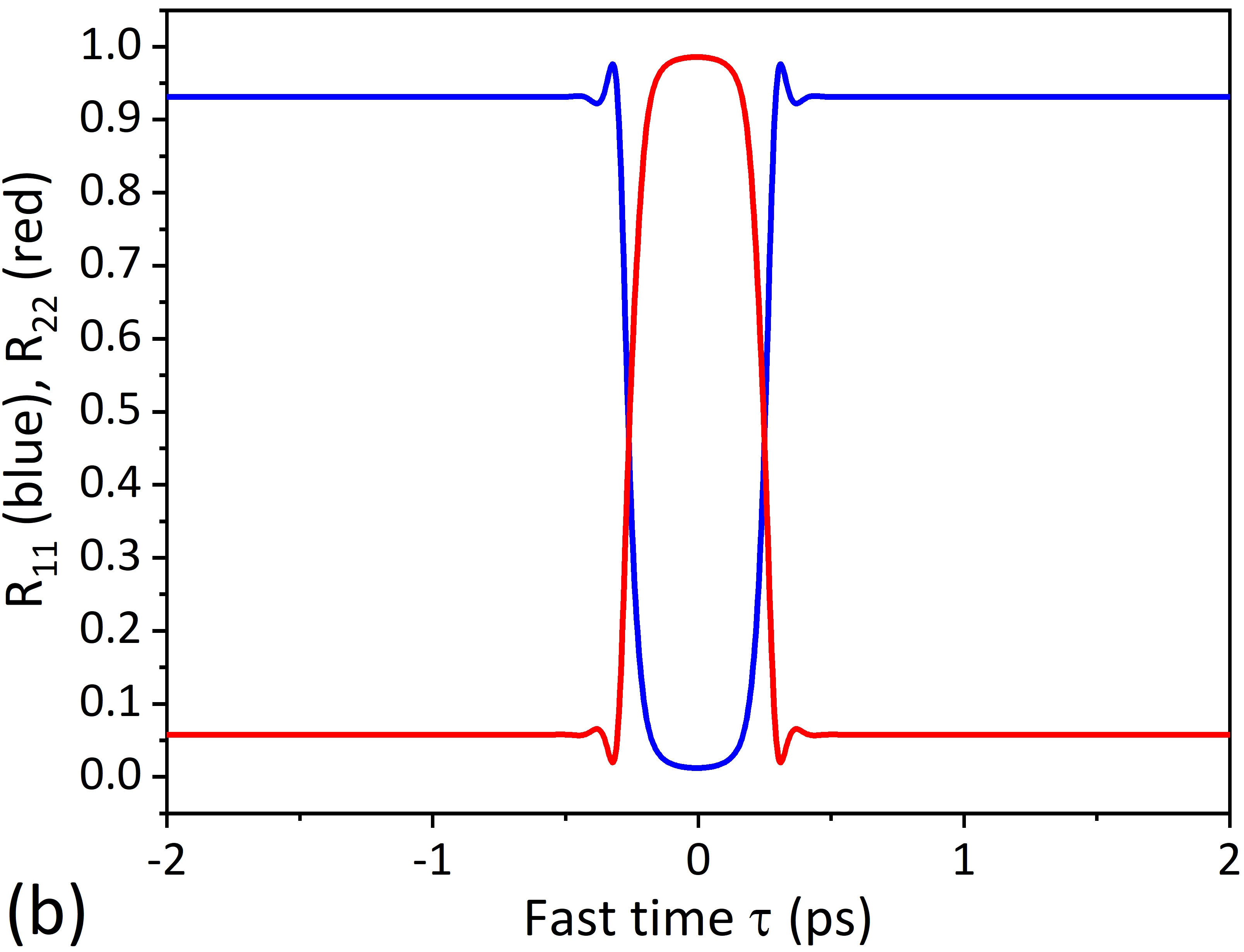}
\caption{a) Intensity profile of a quantum interference TCS for $\theta=-0.5$ ($\theta=-1.0$ in the inset), $|E_2|^2=0.1$, $P=2.5$, $2C=35$, $\Delta=0.66$. According to the chosen free spectral range of around $140$ GHz, the FWHM size of the TCS are about $0.27$ ps and $0.18$ ps for the inset. b) Fast time distributions of occupancy probabilities of level $|1\rangle$, $R_{11}$ (blue line), and level $|2\rangle$, $R_{22}$ (red line).}
\label{fig5}
\end{figure}

DWs are not the only localized structures due to quantum interference in this system. Away from the values of the cavity detuning $\theta_M$ where the DWs maintain a fixed distance, the DWs move away from (towards) each other for $\theta>\theta_M$ ($\theta<\theta_M$) when $\Delta=0.66$ and $|E_2|^2=0.1$. When compared to magnetic DWs in solid state physics of hyperbolic tangent shape, our optical DW displays local fast time oscillations due to group-velocity dispersion instead of diffusion. If the cavity detuning $\theta$ is smaller than $\theta_M$, the two DWs move towards each other and, in view of the large oscillation close to the top of the DWs, they lock and form a TCS without requiring modulational instabilities in a way similar to what was described in \cite{Oppo99} for parametric oscillators. For $\theta=-0.5$, for example, the low-intensity HSS eats in the high-intensity HSS until the DWs lock and form a quantum interference TCS (see Fig. \ref{fig5}a) with an underlying STIRAP process of more than $97\%$ since the probability of occupancy of state $|1\rangle$ is very high in the TCS tails and the probability of occupancy of state $|2\rangle$ is very high at the TCS peak (see Fig. \ref{fig5}b). TCS peak intensities and occupancy probabilities over the interval of TCS existence versus the cavity detuning parameter are reported as black circles in Fig. \ref{fig2}. Typical full-width-at-half-maximum (FWHM) sizes of TCS due to quantum interference are expected to be in the sub-picosecond regimes.
DWs and TCSs formed by locked DWs are localized quantum interference structures robust to perturbations, do not require modulational instabilities and exist in wide ranges of the parameter space of the model equations making them likely to be implemented experimentally. It is important to note that stable DWs and locked DWs have been observed experimentally for two-level media with normal dispersion \cite{Xue15,Garbin17,Gilles17,Garbin21} in regimes not affected by the presence of pattern solutions. The quantum interference DWs and TCSs presented here operate instead in the anomalous dispersion regime.

\begin{figure}[htbp]
\centering
\includegraphics[scale=0.15]{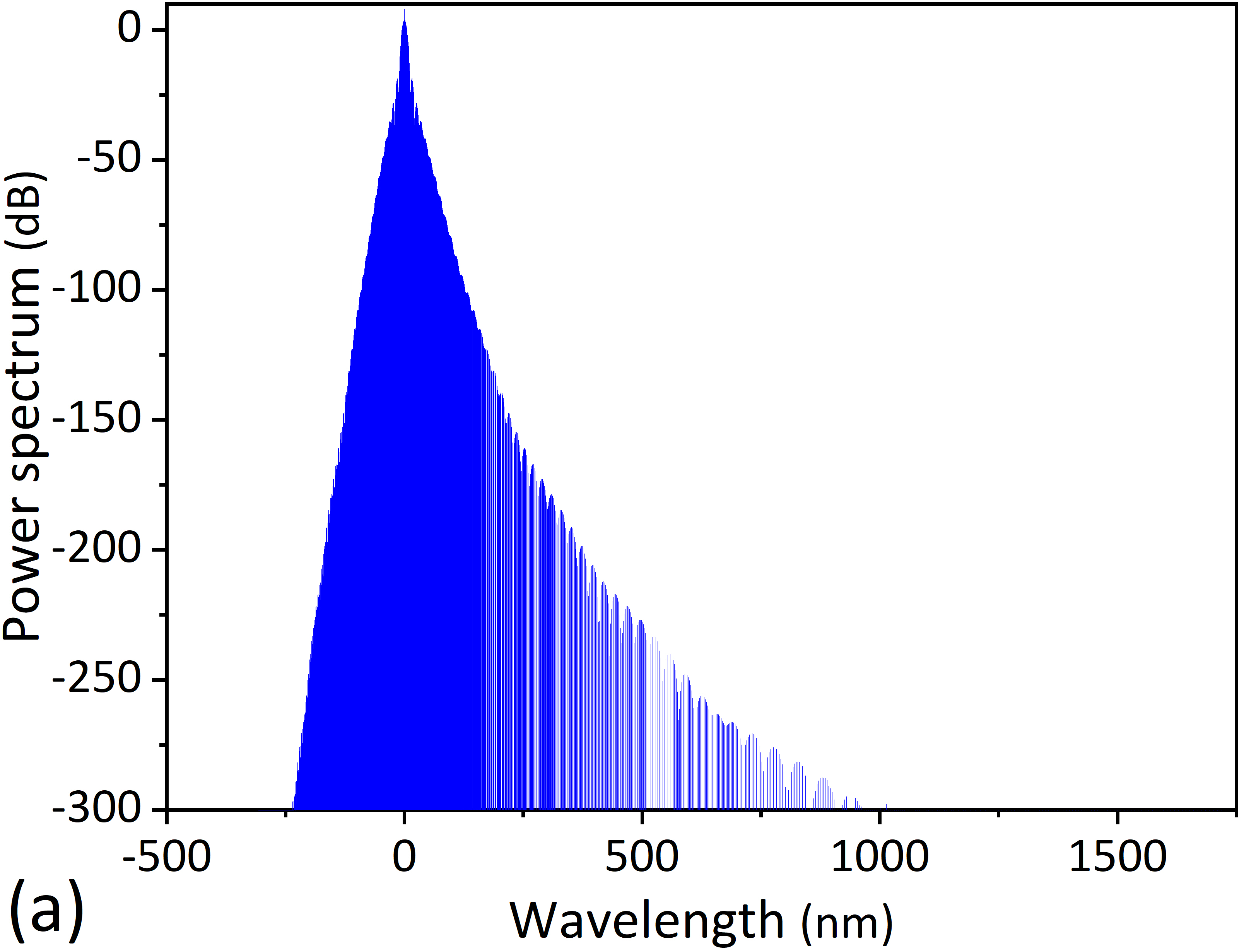}\quad
\includegraphics[scale=0.15]{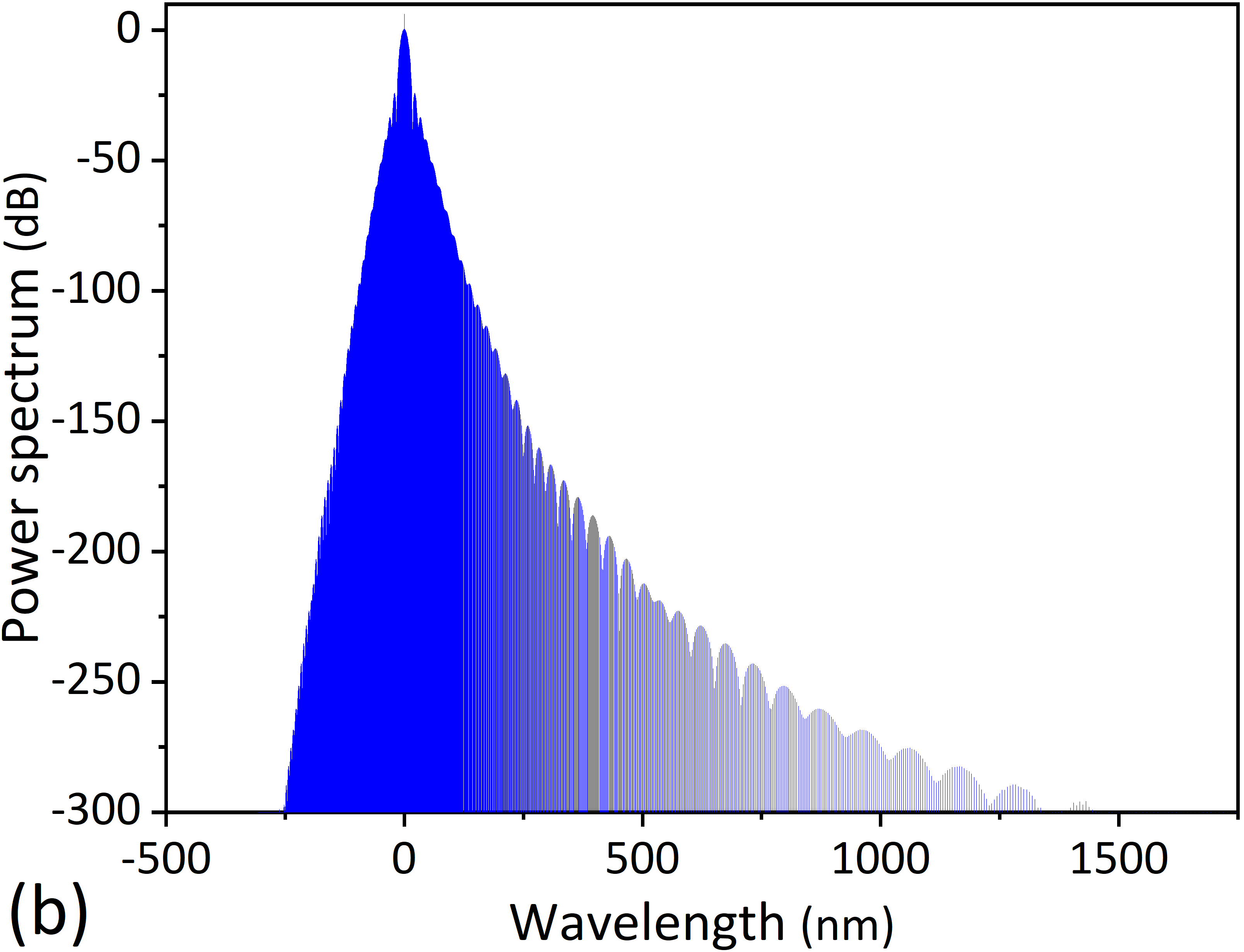}
\caption{Spectrum of the locked DW TCS of Fig. \ref{fig5} for $\theta=-0.5$ and for $\theta=-1.0$. Other parameters in the text.}
\label{fig6}
\end{figure}

Quantum interference TCSs formed by locked DWs have a very high contrast (visibility) since their tails are anchored on the low-intensity branch. There are also almost no local modulations at the bottom of the TCS peaks making them quite different from TCSs in Kerr resonators and excellent candidates for the generation of frequency combs. Fig. \ref{fig6} shows two examples of frequency combs for the values of $\theta=-0.5$ and $\theta=-1.0$. In the first case we have a higher peak intensity TCS with higher definition but narrower spectrum, while in the second case we have a narrower TCS resulting in a broader spectrum. These are excellent realizations of frequency combs made of quantum interference dark states and cw-STIRAP.

Quantum interference between separate two-level transitions leads to CPT, EIT and the population of dark states in three level $\Lambda$ configurations. Here we have shown that quantum interference can also be responsible for the generation of stable DWs and TCSs due to the locking of DWs in microresonators driven by two external fields close to the medium resonances and in the presence of anomalous group velocity dispersion. The DWs separate two dark states inside the optical cavity where the medium is almost exclusively in one or the other of the two ground state energy levels. This phenomenon corresponds to the realization of cavity STIRAP with no input pulses. 

At difference with two-level media, TCSs via quantum interference and anomalous dispersion can occur without modulational instabilities and close to medium and cavity resonances due to EIT, very low absorption and enhanced nonlinear features. Close to resonance TCSs have optimal shapes for nonlinear absorbers owing to their high peak intensities, very low backgrounds and almost no oscillation in the tails. For these reasons, TCSs due to quantum interference can be optimal candidates for the generation of novel microresonator frequency combs with applications in frequency standards, optical communications and high resolution spectroscopy. DWs and TCSs are robust features and are expected for wide ranges of input intensities, medium responses and detunings. Dynamical TCSs, broadening of frequency combs, generalizations to detuned fields $E_2$ of large Rabi frequencies leading to Fano-like resonances as well as quantum interference in $V$ and ladder three-level media in microresonators are presently under investigation.

\end{document}